\documentstyle[11pt,newpasp,twoside,epsf]{article}
\reversemarginpar

\begin{document}
\title{Spectroscopy of the Millennium Outburst and Recent Variability of the Yellow Hypergiant Rho Cassiopeiae}

\author{A. Lobel\altaffilmark{1}, R. P. Stefanik, G. Torres, R. J. Davis} 
\affil{Harvard-Smithsonian Center for Astrophysics, 60 Garden Street, Cambridge 02138 MA, USA}

\author{I. Ilyin}
\affil{Astronomy Division, PO Box 3000, 90014 University of Oulu, Finland}

\author{A. E. Rosenbush}
\affil{Main Astronomical Observatory of the National Academy of Sciences, Ukraine}

\altaffiltext{1}{Guest investigator of the UK Astronomy Data Centre.}

\begin{abstract}

In the summer and fall of 2000 the yellow hypergiant $\rho$ Cassiopeiae
dimmed by more than a visual magnitude, while its effective temperature
decreased from $\sim$7000 K to below 4000 K over $\sim$200 d. 
We observed the highest mass-loss rate of $\sim$5\% of the solar mass
per year in a single stellar eruption so far (Lobel et al. 2003, ApJ, 583, 923).

It is the third outburst of $\rho$ Cas on record in the last century. 
During the outburst the enigmatic cool luminous
hypergiant changed its spectral type form early F- to early M-type. 
The outburst produced an outward propagating circumstellar shock wave, 
resulting in a tremendous cooling of the entire atmosphere.
The optical spectrum became comparable to that of the red supergiant Betelgeuse, 
and revealed strongly blue-shifted molecular absorption bands of titanium-oxide (TiO). 
We determine from the newly formed TiO bands a gas mass-loss rate of 
$\dot{M}$$\simeq$5.4 $\times$$10^{-2}$ $\rm M_{\odot}$~$\rm yr^{-1}$, 
which is of the same order of magnitude as has been proposed for the giant 
outbursts of the Luminous Blue Variable $\eta$ Carinae.

During the pulsation cycles that followed the millennium outburst 
$\rho$ Cas brightened up after mid 2002, and started to dim in early March 2003. 
Over the past two years since the outburst event 
we observe a very prominent inverse P Cygni profile in Balmer H$\alpha$.
Strong H$\alpha$ emission has not before been observed in the hypergiant
over this long period of time, signaling an unusual strong collapse of the 
upper hydrogen atmosphere, which we also observed in the months before the 2000 outburst. 
Since the brightness decrease of March 2003 we observe a remarkable
transformation of the H$\alpha$ profile into a P Cygni profile, signaling 
a strong expansion of Rho Cas' upper atmosphere, which could possibly result in a new
outburst event.

Our radial velocity monitoring shows that the photospheric lines strongly shifted to longer 
wavelengths in the months before January 2003. High-resolution spectroscopic
observations reveal that the lower photosphere rapidly expanded until May-June
2003. Very recent high-resolution observations of September 2003 show however that the
photospheric absorption lines did not shift far blueward as observed in July 2000 before 
the strong brightness decrease of the outburst. The Fe~{\sc i} $\lambda$5572 
is redshifted, signaling the collapse of the lower photosphere. A new strong 
brightness decrease by more than a magnitude in $V$ for the fall and winter of
2003 is therefore not expected.
\end{abstract}

\begin{figure}
\plotfiddle{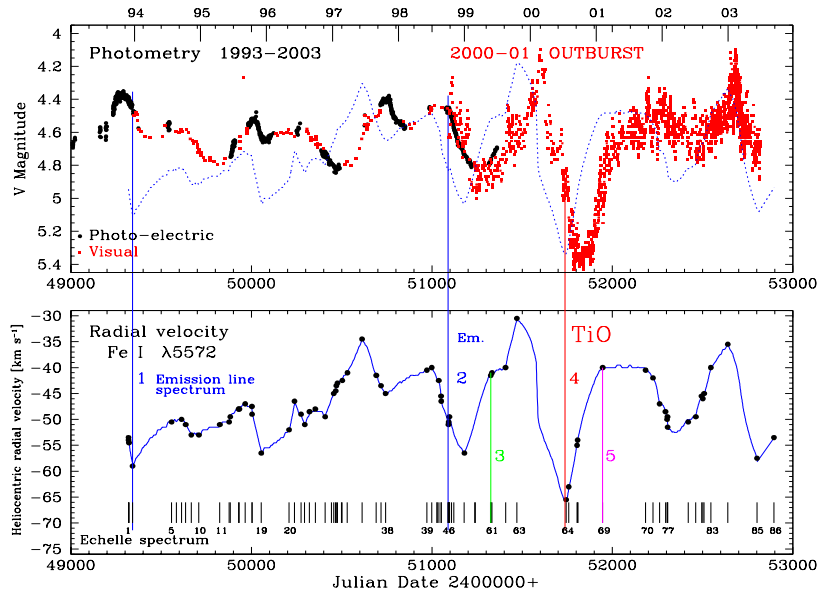}{7cm}{-90}{45}{41}{-190}{230}
\caption{The $V$-brightness curve of $\rho$ Cas is compared in the upper panel to the radial velocity curve,
observed over the last decade. Observation dates of echelle spectra are marked with short vertical lines in 
the lower panel. The radial velocity curve of Fe~{\sc i} $\lambda$5572 
shows a strong increase of the photospheric pulsation amplitude before the outburst of fall 2000 
(JD 2451800$-$JD 2451900), when TiO bands develop ({\em marked TiO}).   }\label{fig-1}
\end{figure}

\section{Introduction}

$\rho$ Cas is a cool hypergiant, one of the most luminous cool massive stars presently
known. Yellow hypergiants are post-red supergiants, rapidly evolving toward
the blue supergiant phase. They are among the prime candidates for progenitors
of Type II supernovae in our Galaxy.  This type of massive supergiant is very important to investigate the
atmospheric dynamics of cool stars and their poorly understood 
wind acceleration mechanisms. These wind driving mechanisms are also important 
to study the physical causes for the luminosity limit of evolved stars
(Lobel 2001). $\rho$ Cas is a rare bright cool hypergiant, which we are
continuously monitoring with high spectral resolution for about a full decade. Its He-core burning phase 
is accompanied by tremendous episodic mass-loss events, which we recently
observed with a new outburst between July 2000 and April 2001. 
(Lobel et al. 2003a).  

In quiescent pulsation phases $\rho$ Cas is a luminous late F-type supergiant
(Lobel et al. 1994).   
During a tremendous outburst of the star in 1945-47 strong absorption bands
of titanium-oxide (TiO) suddenly appeared in the optical spectrum, together with 
many low excitation energy lines, not previously observed.
These absorption lines, normally observed in M-type supergiants,
were strongly blue-shifted, signaling the ejection of a cool circumstellar 
gas shell. In July 2000 we observed the formation of new TiO bands during a strong 
$V$-brightness decrease by $\sim$$1^{\rm m}.4$ ({\it Fig. 1}). 
Our synthetic spectrum calculations show that $T_{\rm eff}$ decreased
by at least 3000 K, from 7250 K to $\simeq$4250~K, and the spectrum became comparable to the early 
M-type supergiants $\mu$~Cep and Betelgeuse. The TiO bands  
reveal the formation of a cool circumstellar gas shell with $T$$<$4000
K due to supersonic expansion of the photosphere and upper atmosphere.
We observe a shell expansion velocity of $v_{\rm exp}$=35$\pm$2~$\rm
km\,s^{-1}$ from the TiO bands. 
From the synthetic spectrum fits to these bands we compute an exceptionally
large mass-loss rate of $\dot{M}$$\simeq$5.4\,$\times$\,$10^{-2}$~$\rm M_{\odot}\,yr^{-1}$,
comparable to the values estimated for the notorious outbursts of $\eta$
Carinae. 

\begin{figure}
\plotone{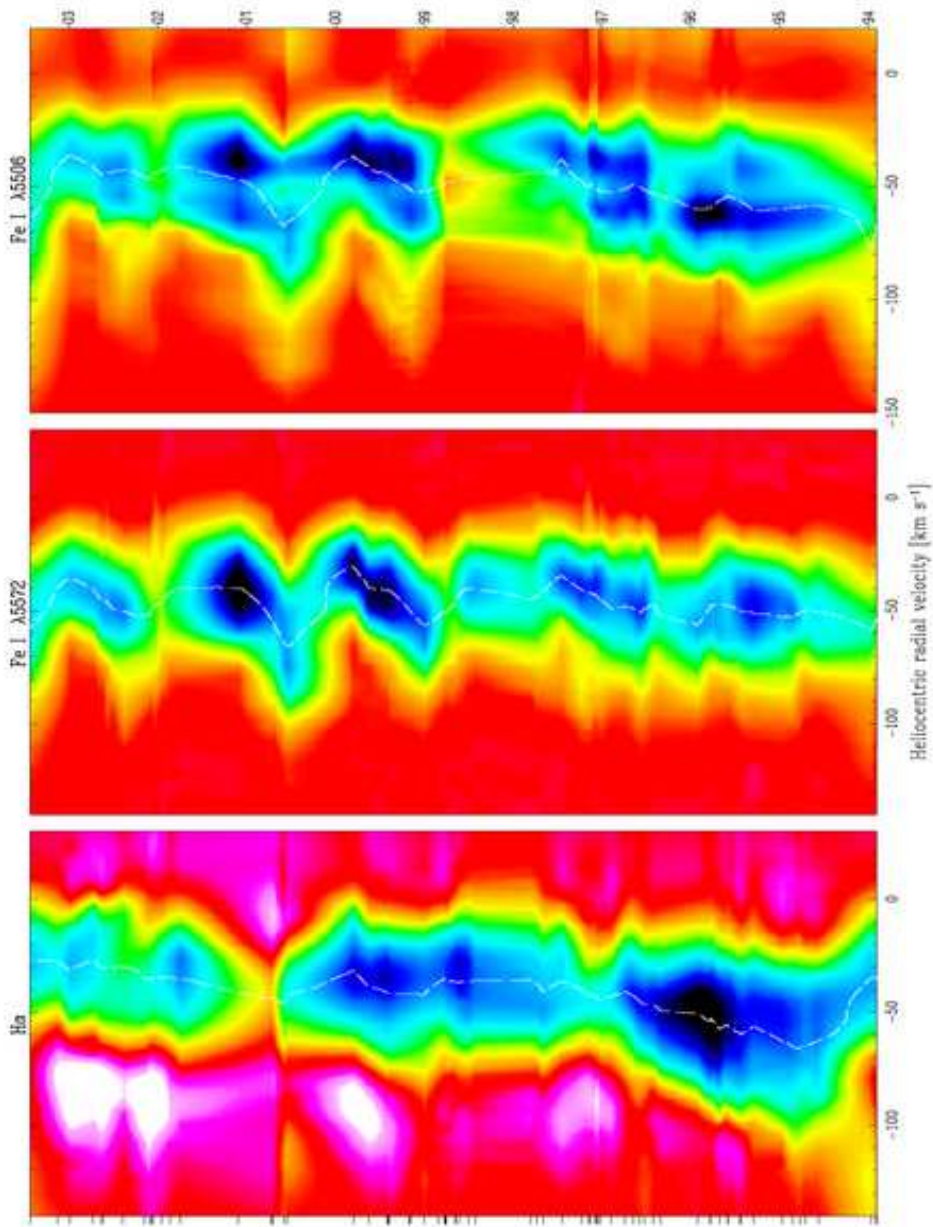}
\caption{Dynamic spectra of H$\alpha$, Fe~{\sc i} $\lambda$5572, and the split
  Fe~{\sc i} $\lambda$5506 line.
The line profiles are linearly interpolated between consecutive observation nights, marked
with the left-hand tickmarks. Time runs upward, indicated for each new calendar year with 
the right-hand numbers. The dashed white lines trace the radial 
velocity determined from the lines.
Notice the strong blue-shift of the Fe~{\sc i} lines during the outburst (mid 2000). 
The outburst is preceded by very strong emission in the short-wavelength wing of H$\alpha$, 
while the absorption core extends longward, and the photospheric Fe~{\sc i} lines strongly red-shift. 
A strong collapse of the entire atmosphere precedes the outburst.}
\end{figure}

\section{Radial Velocity and Brightness Curves}

The upper panel of Figure~1 shows photo-electric observations of $\rho$ Cas 
in the $V$-band ({\it black dots}) over the past decade, 
supplemented with visual magnitude estimates from the $AAVSO$, and the $AFOEV$
(French Association of Variable Star Observers) during the outburst 
of late 2000 ({\it red dots}). The brightness curve shows semi-regular 
variability, with the deep outburst minimum of $V$$\sim$$5^{\rm m}.3$ in
2000 September--November 2000, preceded by a conspicuously bright visual maximum ($V$$\sim$$4^{\rm m}$) 
in March 2000. 

\begin{figure}
\plotfiddle{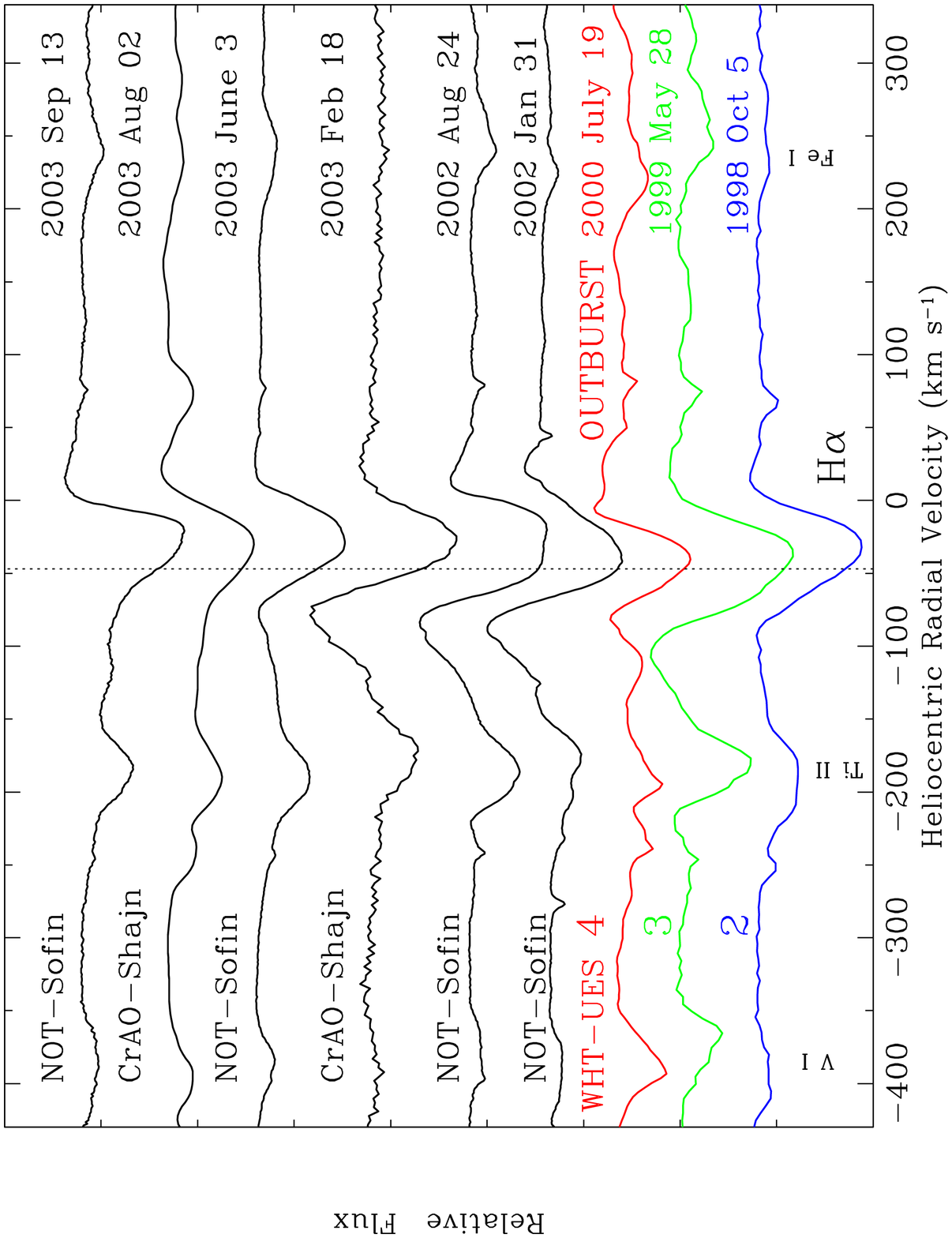}{6.3cm}{-90}{45}{37}{-175}{205}
\caption{The H$\alpha$ absorption core becomes very weak during the outburst.
Unlike the photospheric absorption lines, the H$\alpha$ line core does not blueshift.
Strong emission is observed in the blue wing of H$\alpha$
for the pre-outburst cycle, during the atmospheric collapse that precedes the large
$V$-brightness maximum before the outburst. Strong emission is also observed over
the past two years, while the profile transformed into a P Cygni profile after
the brightness maximum of March 2003.}
\end{figure}

The lower panel shows the radial velocity curve which has been monitored from the 
unblended Fe~{\sc i} $\lambda$5572 absorption line ({\it black dots}).
The radial velocity curve, determined from a linear interpolation of the 
temporal Fe~{\sc i} line profile changes, is compared with the $V$-magnitude 
curve in the upper panel ({\it blue dotted line}). We observe that the star becomes 
brightest for variability phases when the atmosphere rapidly expands. 
$V_{\rm rad}$ decreases by $\sim$20 $\rm km\,s^{-1}$ in less than 200 d during 
the outburst event. The short black vertical lines mark a total of 86
echelle spectra observed with high-resolution spectrographs
over the past 10 years.
The spectra have been obtained from our long-term monitoring campaigns 
with four telescopes in the northern hemisphere;  the Utrecht Echelle Spectrograph 
of the William Herschel Telescope (La Palma, Canary Islands), the Sofin spectrograph
of the Nordic Optical Telescope (La Palma, Canary Islands), the Ritter Observatory
telescope (OH, USA), and the Zeiss-1000 telescope of the Special Astrophysical Observatory 
of the Russian Academy of Science (Zelenchuk, Russia) (Lobel et al. 2003a).  

\section{Dynamic Spectra of 1993--2003} 

Figure 2 shows dynamic spectra of H$\alpha$, Fe~{\sc i} $\lambda$5572, and the
split Fe~{\sc i} $\lambda$5506 line, observed between 1993 and September 2003. 
The white spots in H$\alpha$ are emission above the stellar continuum level.
The radial velocity curves of the H$\alpha$ absorption, and the photospheric Fe~{\sc i} lines 
({\it white dashed lines}) reveal a strongly velocity stratified dynamic
atmosphere (Lobel et al. 1998, Lobel et al., 2003b).
Notice the large blueshift of the Fe~{\sc i} lines during the outburst of mid
2000. The outburst is preceded by very strong line emission in the short wavelength wing of H$\alpha$, 
while the absorption core extends longward (comparable to an inverse P Cygni
profile), and the Fe~{\sc i} lines strongly shift redward. A strong collapse
of the upper H$\alpha$ atmosphere and the deeper photosphere precedes the
outburst event during the pre-outburst cycle of 1999 ({see also \it Fig. 1}).

\section{P Cygni profiles in Balmer H$\alpha$} 

Over the past two years since the outburst we observed an unusually strong
inverse P Cygni profile in H$\alpha$, indicating a new collapse of $\rho$
Cas' extended upper hydrogen atmosphere. During the past five months, after the visual
brightness maximum of March 2003 ({\em Fig. 1}), we observe how the H$\alpha$ line transformed from an 
inverse P Cygni profile into a P Cygni profile ({\it Fig. 3}). The clear P Cygni
line shape for H$\alpha$ is very remarkable because it was not as prominently
observed during the 2000 outburst. It indicates that an even stronger expansion
of the upper atmosphere occurred after early 2003. We presently (September 2003)
measure expansion velocities up to 140~$\rm km\,s^{-1}$ for the H$\alpha$
atmosphere, while the stellar spectral type has changed from mid F to early K, 
based on the recent photospheric spectra. 

\begin{figure}
\plotfiddle{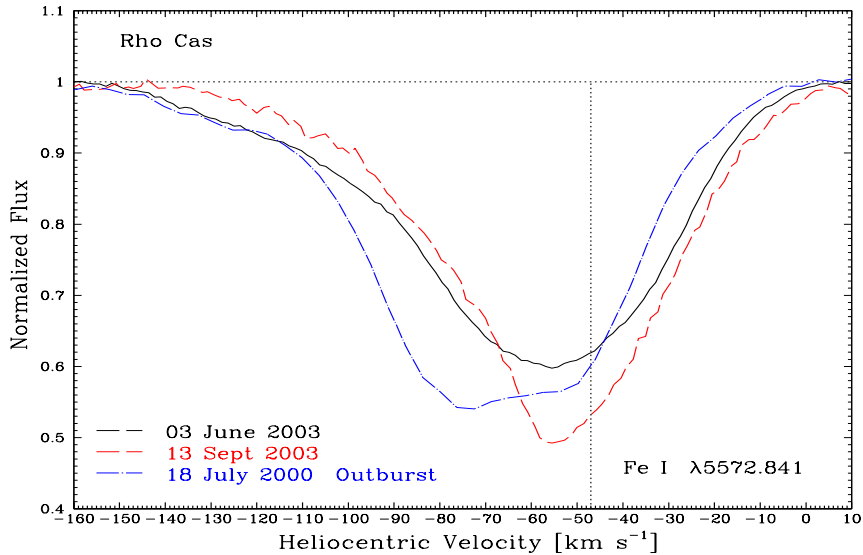}{6.6cm}{-90}{45}{37}{-175}{210}
\caption{High-resolution observations of Fe~{\sc i}
$\lambda$5572 reveal a redshifted absorption line core in September 2003 compared to June
2003. The line profile variations indicate a strong expansion of the
photosphere after early 2003, however not as strong as during the summer of 2000 in
the months preceding the deep brightness minimum of the millennium outburst.}
\end{figure}

\section{Recent Evolution of Photospheric Lines in 2003}

Figure 4 compares high-resolution profiles of the Fe~{\sc i} $\lambda$5572
line, recently observed with NOT-Sofin in June and September 2003, with the detailed line 
shape of July 2000 before the brightness minimum of the outburst. 
After the recent brightness maximum of March 2003 we observed that the core of the 
absorption line strongly shifted toward shorter wavelengths ({\em solid
  drawn line}). The line developed a far violet extended absorption wing, which we also
observed in the months before the outburst event of 2000 ({\em dash-dotted line}). 
The violet line wing signals maximum wind expansion velocities up 
to $-$120 $\rm km\,s^{-1}$ with respect to the stellar rest velocity of $-$47 
$\rm km\,s^{-1}$ ({\em vertical dotted line}). More recent observations of 
September 2003 show however that the core of the photospheric line does not
shift further bluewards ({\em dashed line}), as we observed in July 2000.

From the half-width at half intensity minimum (HWHM) of the line we measure that the radial 
velocity of the photosphere increased from $-$57.5 $\rm km\,s^{-1}$ in June
2003 to $-$53.5 $\rm km\,s^{-1}$ in September 2003. The present collapse
indicates that the expansion of the lower photosphere, after the brightness
maximum of March 2003, was not as strong as during the summer of 2000 when 
the line HWHM assumed $-$65.5 $\rm km\,s^{-1}$. The visual brightness of
$\rho$ Cas decreased by about half a magnitude since March 2003. The 
recent spectroscopic observations indicate however that a possible further dimming to a new 
deep brightness minimum is not expected for the fall and winter of 2003.  
Our continuous monitoring programs remain active to further document and investigate the 
spectral variability of this enigmatic cool hypergiant. 

\section{Conclusions}

We observe the formation of TiO bands in the spectrum of $\rho$ Cas before the deep brightness minimum
in the outburst of 2000-01. A supersonic expansion velocity is observed for the new TiO bands, while $V$ 
rapidly dims by more than a full magnitude.  A large oscillation cycle and a
large brightness maximum, with $T_{\rm eff}$ above 7250 K, precede the outburst minimum.
We compute that $\dot{M}$$\simeq$5.4\,$\times$\,$10^{-2}$ $M_{\odot}$\,$\rm yr^{-1}$ 
during the event, whereby $T_{\rm eff}$ decreases from $\simeq$7250 to $<$3750 K. 
$T_{\rm eff}$ returns to 5750 K within 100 d after the deep outburst minimum.
Since recurrent outbursts occur about every half century in $\rho$ Cas, these outburst phases 
of punctuated enhanced mass-loss are the major mass-loss mechanism of this massive cool hypergiant.

\acknowledgments
This research is based in part on data obtained with the William Herschel Telescope
of the ING at La Palma, Canary Islands. Financial support has been provided in 
part by a NASA grant to the Smithsonian Astrophysical Observatory, MA.


\begin{references}
\reference Lobel, A. 2001, \apj, 558, 780
\reference Lobel, A., de Jager, C., Nieuwenhuijzen, H., Smolinski, J., \& Gesicki, K. 1994 \aap, 291, 226
\reference Lobel, A., Dupree, A. K., Stefanik, R. P., Torres, G., Israelian,
G., Morrison, N., de Jager, C., Nieuwenhuijzen, H., Ilyin, I., \& Musaev, F. 2003a, \apj, 583, 923
\reference Lobel, A., Dupree, A. K., Stefanik, R. P., Torres, G., Israelian,
G., Morrison, N., de Jager, C., Nieuwenhuijzen, H., Ilyin, I., \& Musaev,
F. 2003b, Proceedings of IAU Symposium No. 210, {\em Modelling of Stellar Atmospheres}, 17 - 21 June 2002, Uppsala, Sweden. To appear in ASP Conf. Ser. 2003, N. E. Piskunov, W. W. Weiss, D. F. Gray, eds.
\reference Lobel, A., Israelian, G., de Jager, C., Musaev, F., Parker, J. Wm., \& Mavrogiorgou, A. 1998, \aap, 330, 659
\end{references}
\end{document}